\documentclass[prb,twocolumn,showpacs,amsmath,amssymb,superscriptaddress,floatfix]{revtex4}

\usepackage{ graphicx } 
\usepackage{ epstopdf }
\usepackage{ bm } 
\usepackage{ color }

\newcommand{\be}{\begin{equation}}
\newcommand{\ee}{\end{equation}}
\newcommand{\bea}{\begin{eqnarray}}
\newcommand{\eea}{\end{eqnarray}}

\newcommand{\ket}[1]{\left | #1 \right\rangle}
\newcommand{\bra}[1]{\left\langle #1 \right|}

\begin{document}
\title{Reduced density matrix and order parameters of a topological insulator}
\author{ Wing Chi Yu }
\affiliation{ \textit Department of Physics and ITP,
The Chinese University of Hong Kong, Hong Kong, China}
\author{ Yan Chao Li }
\affiliation{ \textit College of Materials Science and
Opto-Electronic Technology, University of Chinese Academy of
Sciences, Beijing, China }
\author{ P. D. Sacramento }
\affiliation{ \textit CeFEMA,
Instituto Superior T\'ecnico, Universidade de Lisboa, Av. Rovisco Pais, 1049-001 Lisboa, Portugal and
Beijing Computational Science Research Center, China }
\author{ Hai-Qing Lin }
\affiliation{ \textit
Beijing Computational Science Research Center, Beijing, China
}

\date{ \today }

\begin{abstract}
It has been recently proposed that the reduced density matrix may be
used to derive the order parameter of a condensed matter system.
Here we propose order parameters for the phases of a topological
insulator, specifically a spinless Su-Schrieffer-Heeger (SSH) model, and consider the
effect of short-range interactions. All the derived order parameters
and their possible corresponding quantum phases are verified by the
entanglement entropy and electronic configuration analysis results. The order
parameter appropriate to the topological regions is further proved by
calculating the Berry phase under twisted boundary conditions. It is
found that the topological non-trivial phase is robust to the
introduction of repulsive inter-site interactions, and can appear
in the topological trivial parameter region when appropriate
interactions are added.

\end{abstract}
\pacs{05.30.Rt, 03.67.-a, 71.10.Fd}


\maketitle

\section{Introduction}

At absolute zero temperature, a quantum many-body system can undergo
a quantum phase transition (QPT) \cite{Sachdev2000,Carr2011} by
varying a non-thermal external driving parameter such as the
magnetic field. Across the quantum critical point, the qualitative
structure of the many-body ground state wavefunction undergoes a
significant change and the change is completely driven by the
quantum fluctuation in the ground state. To characterize a
continuous quantum phase transition, a traditional approach is to
use Landau's symmetry breaking theory in which the order parameter
plays the central role. The order parameter is nonzero in the
symmetry broken phase while it vanishes in other phases. Through the
emergency of the order parameter, the phase boundary can also be
determined. However, to find an appropriate order parameter describing
certain phase is a non-trivial task. People have to rely on physical
intuition or resort to methods such as group theory and the
renormalization group analysis. A prior knowledge of the symmetry
breaking of the system is required and the methods are not always
guaranteed to apply, especially to systems exhibiting
topological QPTs \cite{Wen2004}.

On the other hand, in the recent decade much attention has been paid to investigate quantum phase
transitions from the perspective of quantum information science.
One of the examples is the study of quantum entanglement in quantum critical
phenomena \cite{Amico2008,Osborne2002,Osterloh2002}. Being a measure of quantum
correlation, it is believed that the entanglement plays a crucial role in QPTs.
Studies showed that the quantum entanglement helps to witness quantum critical
points and exhibits interesting properties such as scaling \cite{Osborne2002,Osterloh2002},
singularity or maximum \cite{GuEE}, etc., in various transitions. It was also shown to
be capable of detecting topological orders \cite{Levin2006,Kitaev2006}. In contrast to
the traditional approach, the application of the quantum entanglement does not require a
prior knowledge of the system's symmetry and this makes it a great advantage to use
for the study of QPTs.

Recently, along the streamline of quantum entanglement, Gu, Yu and
Lin \cite{Gu2013} proposed a systematic way to derive the order
parameter by studying mutual information and the spectra of the
corresponding reduced density matrix. To apply the scheme, one only
needs the knowledge of the ground state of the system but not the
symmetries existing in it. By studying the single site and two sites
reduced density matrices, the order parameters for the spin-density
wave (SDW), charge-density wave (CDW), bond-order wave (BOW) and the
phase separation phase (PS) in the one-dimensional extended Hubbard
model were successfully obtained \cite{YU2016}. Meanwhile, there are
other independent proposals to derive the order parameter. Furukawa,
Misguich, and Oshikawa \cite{Furukawa2006} proposed a variational
method by investigating a set of low-energy “quasi-degenerate”
states that lead to the symmetry breaking in the thermodynamic
limit. Their scheme was later improved by Henley and Changlani
\cite{Henley2014}. Cheong and Henley \cite{Cheong2009} on the other
hand suggested to study the singular-value decomposition of the
correlation density matrix to gain information on the correlation
function and the order parameter. Compared to those methods, the one
proposed by Gu \emph{et al.}\cite{Gu2013} is a non-variational
approach and is relatively more intuitive to apply. Moreover, it
also establishes a connection between the mutual information and the
order parameters.

In this work, we apply Gu \emph{et al.}'s method to a problem which has topological properties.
The model considered is a one-dimensional spinless fermions Su-Schrieffer-Heeger (SSH)-like model
with explicit dimerization.
The original SSH model \cite{ssh} describes the coupling between spinfull electrons
and phonons and was proposed to describe the one-dimensional conducting polyacetylene;
the condensation of the phonons leads to a dimerization of the lattice.
The simplified model considered here, with explicit dimerization, can be viewed
as a two-band model where interband hopping with alternating amplitudes
takes place at the same site or neighboring sites. The model has two
gapped phases, depending on the relative amplitudes of the two sets of hoppings.
If the two sets of hoppings are equal (no dimerization), the spectrum is gapless.
One of the phases
is topologically trivial while the other has a nonvanishing winding number
and fermionic edge states. The model has no true topological order but is a
symmetry-protected topological system \cite{hasan,szhang,schnyder}.
The model is also related to Shockley's model (see
for instance \cite{yakovenko}).
We identify the order parameters appropriate to describe the two phases: in the trivial
phase the order parameter is fully local and involves the two bands at a given site.
In the topological phase the order parameter involves two neighboring lattice sites.

The effect of interactions is also addressed. Both the separate
addition of dimerization and interactions lead to spectra that are
gapped. In the case of spinless fermions Pauli's principle forbids a
Hubbard-like term and a nearest-neighbor interaction term is the
dominant. Dimerization and the consequent existence of two bands
allows a local interband Hubbard term. These various cases have been
extensively studied before using various techniques such as
bosonization and the density matrix renormalization group method
(DMRG)\cite{white}. The addition of interactions to the problem leads to a
competition between various orderings. In the case of spinfull
systems there is a competition between bond-ordered, charge density
waves and spin density waves \cite{wang,jeckelmann,riera,zhang1} as
a result of the phonon and electronic repulsion terms. The
competition in the case of explicit dimerization has also been
addressed \cite{benthien} as well as the contribution of bond-bond
and mixed bond-site electronic couplings \cite{campbell}. These
competitions continue to attract interest in the literature (see for
instance, \cite{ejima,kumar,weber}). The non-dimerized problem but
with interactions has also been shown to lead to spontaneous
dimerization as a result of interactions in a narrow region between
the ordered SDW and CDW phases
\cite{nakamura,sengupta,zhang2,ejima2}. In this work, we consider the effect of interactions in the spinless fermions using
both DMRG and exactly diagonalizaiton (ED) methods.

One of the goals of this work is to gauge the effectiveness of the method to
construct order parameters using the method described in the next section.
As an essential part it requires calculating the reduced density matrix of a subsystem
of some minimal size (discussed ahead). One may expect some difficulties associated
with a topological system. As shown ahead, the method provides in a direct way a form
for the order parameter in the trivial region but in the topological region some
ambiguity is left due to the significant contribution of all the eigenvalues of the
reduced density matrix. An appropriate change of basis reduces the
number of eigenvalues that contribute significantly. This change of basis is obtained
representing the Hamiltonian in a Majorana fermion basis.

\begin{figure}
\includegraphics[width=0.6\columnwidth]{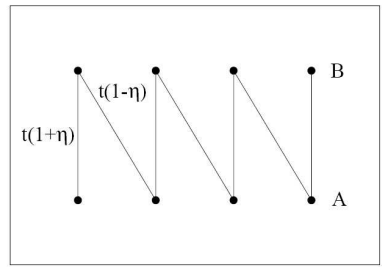}
\caption{\label{fig1}
Two-band SSH model of spinless fermions: at each lattice site there
are two inequivalent sites, $A$ and $B$, linked by alternating
hoppings given by $t(1+\eta)$ and $t(1-\eta)$.
}
\end{figure}

The order parameters for various phases in the interacting system are derived, and are compared to other order parameters, such as the bond-order and charge density wave ones. The topology of the system is affected by the interactions and we use Berry phase
to separate the trivial from the topological regions. Interestingly, the derived order parameter appropriate for the topological regions is robust to the presence of inter-site repulsive interactions.

The paper is organized as follows. In Sec.~\ref{sec:level2}, we
first briefly introduce the scheme to derive the potential
order parameters. Then an introduction about the spinless SSH model
is given in Sec.~\ref{sec:level3}. The topological phase transition
in the model is detected by the entanglement entropy and the
order parameters for the topologically trivial and non-trivial
phases are derived in Sec.~\ref{sec:level4}. In
Sec.~\ref{sec:level5}, we consider the case when interactions
are added. The ground state phase diagram and order parameters for each quantum
phases are obtained. The order parameter corresponds to the topological
non-trivial phase is further verified by the berry phase results.
Finally, a conclusion is given in Sec.~\ref{sec:level6}.

\section{\label{sec:level2} Outline of the scheme in deriving the potential order operators}

To derive the order parameter, we first have to determine the
minimum size of the block (sub-system) for which the mutual
information (also known as the correlation entropy) does not vanish at a long distance.
The mutual information is defined as%
\begin{equation}
S(i,j)=S\left( {\rho }_{i}\right) +S\left( {\rho }_{j}\right) -S\left( {\rho
}_{i\cup j}\right) ,
\label{eq:MI}
\end{equation}%
where \bea S\left( {\rho }_{i}\right) =-tr({\rho }_{i}\ln{\rho
}_{i}) \label{eq:vnS} \eea is the von-Neumann entropy of the block
$i$. $\rho_i$ is the reduced density matrix obtained by tracing out
all other degrees of freedom except those of the block $i$,
i.e.${\rho }_{i}=$tr$\left\vert \Psi _{0}\right\rangle \left\langle
\Psi_{0}\right\vert $ where $\ket{\Psi_0}$ is the ground state of
the system. If and only if the mutual information is non-vanishing
at a long distance, there exists a long-range order (or quasi long range order) in the
system \cite{MMWolf,SJGuJPA}.

The next step is to calculate the eigenvalues and eigenvectors of
the reduced density matrices of the desired block size. Depending on
the basis of the reduced density matrix, it is possible to have
diagonal and off-diagonal long-range orders. In terms of the
creation (annihilation) operator $a_{i\mu}^{\dagger} (a_{i\mu})$ for
a state $\ket{\mu}$ localized at the block $i$, define the diagonal
order operator as \cite{Gu2013}
\begin{eqnarray}
O_i^d=\sum_{\mu\leq\xi}w_{\mu}a_{i\mu}^{\dagger}a_{i\mu},
\label{eq:O_d}
\end{eqnarray}
where $\xi$ is the rank of $\rho_i$. It can be proved that for any
$\mu>\xi$, the operator $a_{i\mu}^{\dagger}a_{i\mu}$ does not
correlate. The coefficients $w_{\mu}$ can be fixed by the traceless
condition tr$(\rho_iO_i^d)=0$ and the cut-off condition
$\max(\{w_{\mu}\})=1$.

If the two-block reduced density matrix $\rho_{i\cup j}$ is not
diagonal in the eigen-basis of $\rho_i\otimes\rho_j$, there exists
off-diagonal long-range order in the system. The corresponding order
operator is defined by
\begin{eqnarray}
O_i^o=\sum_{<\mu,\nu>}w_{\mu\nu}a_{i\mu}^{\dagger}a_{i\nu}+w^*_{\mu\nu}a_{i\nu}^{\dagger}a_{i\mu},
\label{eq:O_o}
\end{eqnarray}
where $\mu\ne\nu$ and the sum is over all the pairs of $\mu, \nu$ that correspond to the
non-zero off-diagonal matrix elements in $\rho_{i\cup j}$.

\section{\label{sec:level3} Spinless SSH model}

\begin{figure}
\includegraphics[width=0.75\columnwidth]{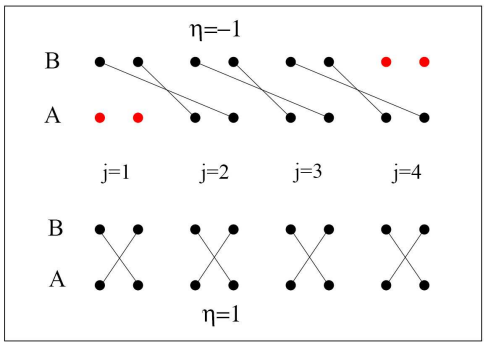}
\caption{\label{fig2}
(Color online)
Phases of SSH model (or Schockley model).
For negative
$\eta$ the model is topologically non-trivial with edge states represented
by the decoupled Majorana operators (each Majorana is represented by a dot). Since at each
end site there are two decoupled Majoranas, these combine to form edge fermionic
modes.
There is also a trivial phase with no zero energy modes
for positive $\eta$.
}
\end{figure}

This model describes a dimerized chain of spinless fermions hopping in a tight-binding band.
The dimerization is parametrized by $\eta$. Due to the dimerization the unit cell
contains two atoms of types $A$ and $B$. The sites are indexed by $j$. The model is given by the Hamiltonian
\bea
H = -\mu & \sum_j & \left(n_{j,A}+n_{j,B} \right)
\nonumber \\
-t & \sum_j &  \left[  (1+\eta) c_{j,B}^{\dagger} c_{j,A} + (1+\eta) c_{j,A}^{\dagger} c_{j,B} \right.
\nonumber \\
&+& \left. (1-\eta) c_{j+1,A}^{\dagger} c_{j,B} +(1-\eta) c_{j,B}^{\dagger} c_{j+1,A} \right].
\nonumber \\
& &
\eea
The operator $c_{j,\alpha}$ destroys a spinless fermion at site $j$ of type $\alpha=A,B$, and $n_{j,\alpha}=c_{j,\alpha}^{\dagger} c_{j,\alpha}$.
The amplitude $t$ is the hopping, $\eta$ is the dimerization and $\mu$ is the chemical potential.
The model is related to the Schockley model \cite{yakovenko} by
taking $t_1=t(1+\eta)$ and $t_2=t(1-\eta)$. The region of $\eta>0$ corresponds to $t_1>t_2$
and vice-versa for $\eta<0$.
The Hamiltonian in real space mixes nearest-neighbor sites and also has
local terms. The links involved are depicted in Fig. \ref{fig1}.

We may define hermitian Majorana operators, $\gamma_{j,\alpha,\beta}$ (with $\beta=1,2$), as
\bea
c_{j,A} &=& \frac{1}{2} \left( \gamma_{j,A,1} + i \gamma_{j,A,2} \right),\nonumber \\
c_{j,B} &=& \frac{1}{2} \left( \gamma_{j,B,1} + i \gamma_{j,B,2} \right).
\eea
In terms of Majorana operators the Hamiltonian is written as
\bea
H &=& -\frac{\mu}{2} \sum_{j=1}^N \left( 2+i\gamma_{j,A,1} \gamma_{j,A,2}
+ i \gamma_{j,B,1} \gamma_{j,B,2} \right) \nonumber \\
&-& \frac{it}{2} (1+\eta) \sum_{j=1}^N \left(
\gamma_{j,B,1} \gamma_{j,A,2} + \gamma_{j,A,1} \gamma_{j,B,2} \right) \nonumber \\
&-& \frac{it}{2} (1-\eta) \sum_{j=1}^{N-1} \left(
\gamma_{j+1,A,1} \gamma_{j,B,2} + \gamma_{j,B,1} \gamma_{j+1,A,2} \right) \nonumber \\
& &
\eea
under open boundary condition. Taking $\mu=0$ we have a couple of special points: i) At $\eta=-1$
we have a state with two
fermionic-like zero energy edge states, since the four operators
$\gamma_{1,A,1}, \gamma_{1,A,2}; \gamma_{N,B,1}, \gamma_{N,B,2}$ are missing from
the Hamiltonian.
ii) An example of a trivial phase is the point $\eta=1$ in which case there
are no zero energy edge states.
In Fig. \ref{fig2} the phases with edge modes are presented for special points
in parameter space.
The model has simplified time-reversal symmetry and sublattice symmetry, if
the chemical potential vanishes. The model is in class BDI and therefore
allows the presence of a $Z$ index related to the winding number and the number
of edge modes.

\begin{figure}
\includegraphics[width=0.9\columnwidth]{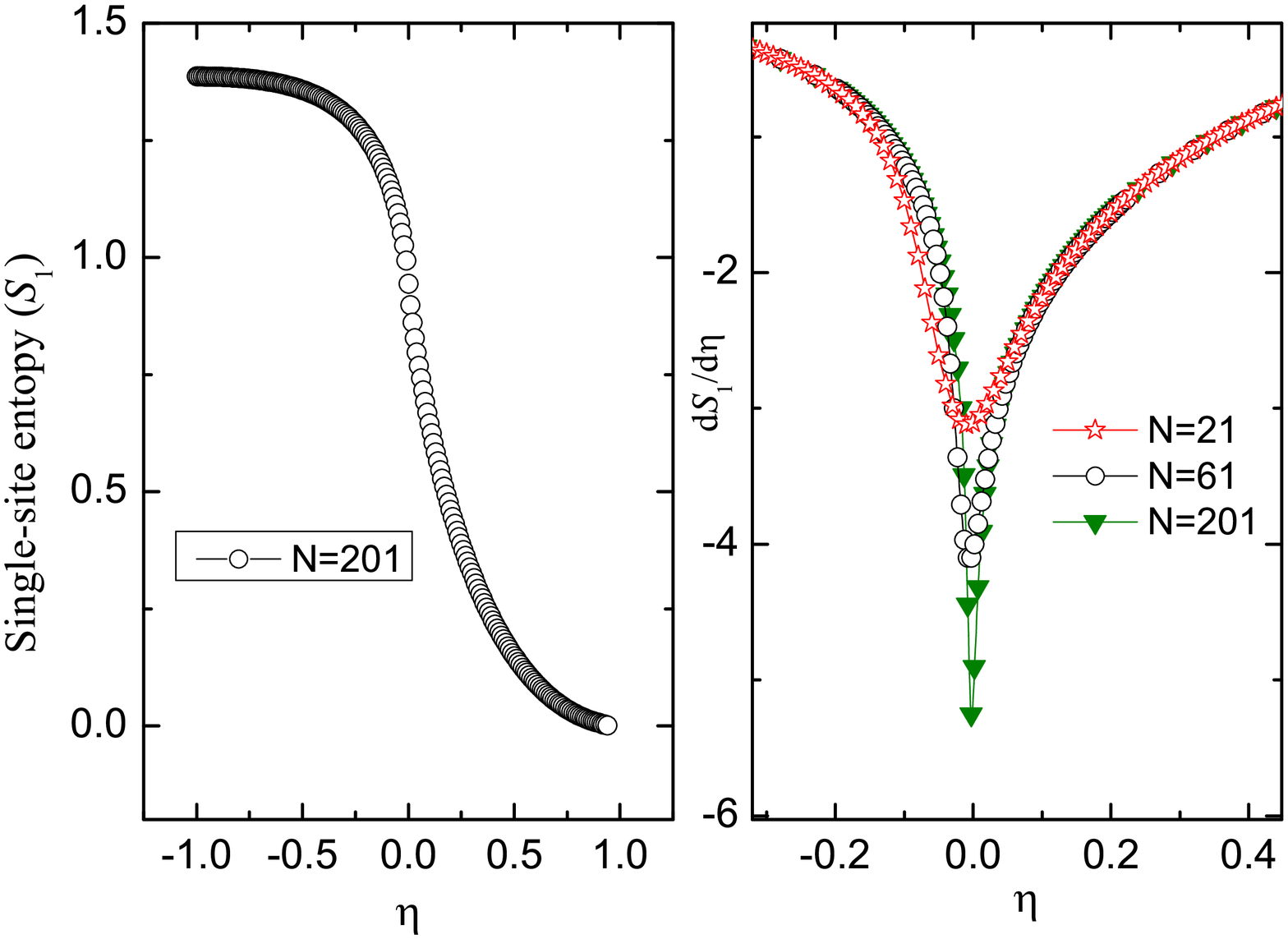}
\caption{\label{fig3} (Color online) Entanglement entropy and its
derivative as a function of dimerization. }
\end{figure}

At the special point of interest $\mu=0,\eta=-1$ shown in the
figure, the Hamiltonian reduces to
\be H = it \sum_{j=1}^{N-1}
\left( \gamma_{j,B,2} \gamma_{j+1,A,1} - \gamma_{j,B,1}
\gamma_{j+1,A,2} \right).
\ee
Let us define non-local fermionic
operators~\cite{kitaev} \bea d_j &=& \frac{1}{2} \left(
\gamma_{j,B,2} + i \gamma_{j+1,A,1} \right),
\nonumber \\
d_j^{\dagger} &=& \frac{1}{2} \left( \gamma_{j,B,2} - i \gamma_{j+1,A,1} \right),
\eea
and
\bea
f_j &=& \frac{1}{2} \left( \gamma_{j,B,1} - i \gamma_{j+1,A,2} \right),
\nonumber \\
f_j^{\dagger} &=& \frac{1}{2} \left( \gamma_{j,B,1} + i \gamma_{j+1,A,2} \right).
\eea
We can show that
\bea
i \gamma_{j,B,2} \gamma_{j+1,A,1} &=& 2 d_j^{\dagger} d_j -1, \nonumber \\
-i \gamma_{j,B,1} \gamma_{j+1,A,2} &=& 2 f_j^{\dagger} f_j -1.
\eea
In terms of these new operators we can write that
\be
H=t \sum_{j=1}^{N-1} \left( 2 d_j^{\dagger} d_j -1 +2 f_j^{\dagger} f_j -1 \right)
\ee
and, therefore, the problem is diagonalized.
It is now clear that the ground state is obtained by taking
$d_j^{\dagger} d_j=0$ and $f_j^{\dagger} f_j=0$  at each site.
This new Hamiltonian in terms of the $d$ and $f$ operators is like
an Hamiltonian with no hopping and just a chemical potential
$\tilde{\mu}=-2t$.

Note that the new operators can be related to the original ones in
terms of a non-local transformation as
\bea
d_j &=& \frac{i}{2}
\left( c_{j,B}^{\dagger} -c_{j,B} + c_{j+1,A} + c_{j+1,A}^{\dagger}
\right),\nonumber \\
f_j &=& \frac{1}{2} \left( c_{j,B}^{\dagger} +c_{j,B} - c_{j+1,A} + c_{j+1,A}^{\dagger} \right).
\label{eq:df}
\eea
Also
\bea
c_{j,A} &=& \frac{1}{2} \left[ -i(-d_{j-1}^{\dagger} +d_{j-1})-(f_{j-1} - f_{j-1}^{\dagger}) \right],
\nonumber \\
c_{j,B} &=& \frac{1}{2} \left[ f_{j}^{\dagger} +f_{j} +i( d_{j} + d_{j}^{\dagger}) \right].
\label{eq:cdf}
\eea

At the special point we are considering we may also write
\be
H = -2t \sum_j \left( c_{j+1,A}^{\dagger} c_{j,B} + c_{j,B}^{\dagger} c_{j+1,A} \right).
\ee

\begin{figure}
\includegraphics[width=0.9\columnwidth]{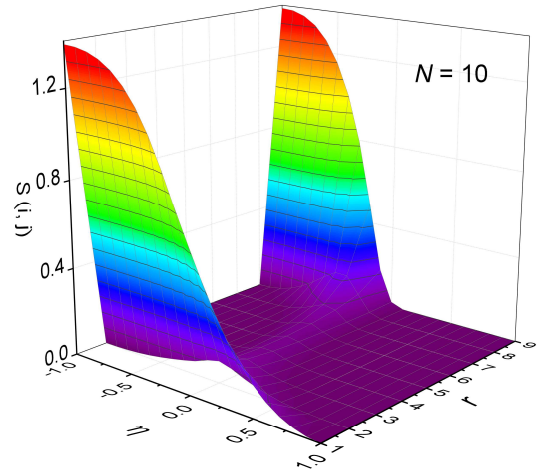}
\caption{\label{fig4} (Color online) Mutual information (correlation
entropy) for $N=10$ using OBC. The sub-system is taken as a
single-site consisting of two atoms of type A and B. }
\end{figure}

\section{\label{sec:level4}Topological insulator}

The evidence that the SSH model has a topologically non-trivial
phase can be provided as above, solving the problem in a finite
chain using open boundary conditions and showing that there are zero
energy edge modes, as shown in Fig. \ref{fig2}. Using the bulk-edge
correspondence it can be shown that the winding number is
non-trivial in the same phase. Methods inspired by quantum
information theory may also be used, such as the entanglement entropy,
and is discussed next.

\subsection{Entanglement entropy}

The entanglement entropy between a single site and the rest of the
chain $S_1$ is defined by the von-Neumann entropy in Eq.~(\ref{eq:vnS}).
As shown in Fig.~\ref{fig3}, it detects the topological phase
transition at $\eta=0$ between the trivial phase and the topological
phase. The transition point is particularly visible if one
calculates the derivative of the entanglement entropy; It becomes
sharper as the system size grows. Even though there
is no change of symmetry as one crosses the gapless point, the
correlations change and this is detected by the entanglement
entropy. In addition, we note that $S_1=0$ at $\eta=1$ and
$S_1=2\ln2$ at $\eta=0$. This may be explained as follows: When $\eta=1$, the inter-site hopping terms equal to zero. There is no information exchange between different sites. Therefore, $S_1$ which measures the entanglement between an arbitrary site and the rest of the chain becomes zero. While for $\eta=0$, the inter-site and intra-site hopping becomes the same. The entanglement for a single site thus reaches its maximum, i.e. $S_1=4\times(-\frac{1}{4}\ln\frac{1}{4})=2\ln2$.

\subsection{Mutual information}

In order to implement the method discussed in Sec \ref{sec:level2}, we calculate the
mutual information or entropy correlation defined in
Eq.~(\ref{eq:MI}) in a system with open boundary condition. The
sub-system is taken as a single-site consisting of two atoms of type
$A$ and $B$. In Fig. \ref{fig4}, $r=|i-j|$ is the distance between
sites $i$ and $j$.

For $\eta >0$, the correlation entropy is vanishing exponentially as $r$ grows. For $\eta<0$, there exists correlation between the two ends of the chain indicative
of the existence of the edge modes. In a finite system they are coupled and their
degeneracy is lifted. In the thermodynamic limit the edge modes become completely decoupled. Around the critical point, $\eta=0$, the correlations extend along the system, signalling the quantum phase transition.

\subsection{Single-site reduced density matrix and order parameters}
\label{sec:4C}

\begin{figure}
\includegraphics[width=1.0\columnwidth]{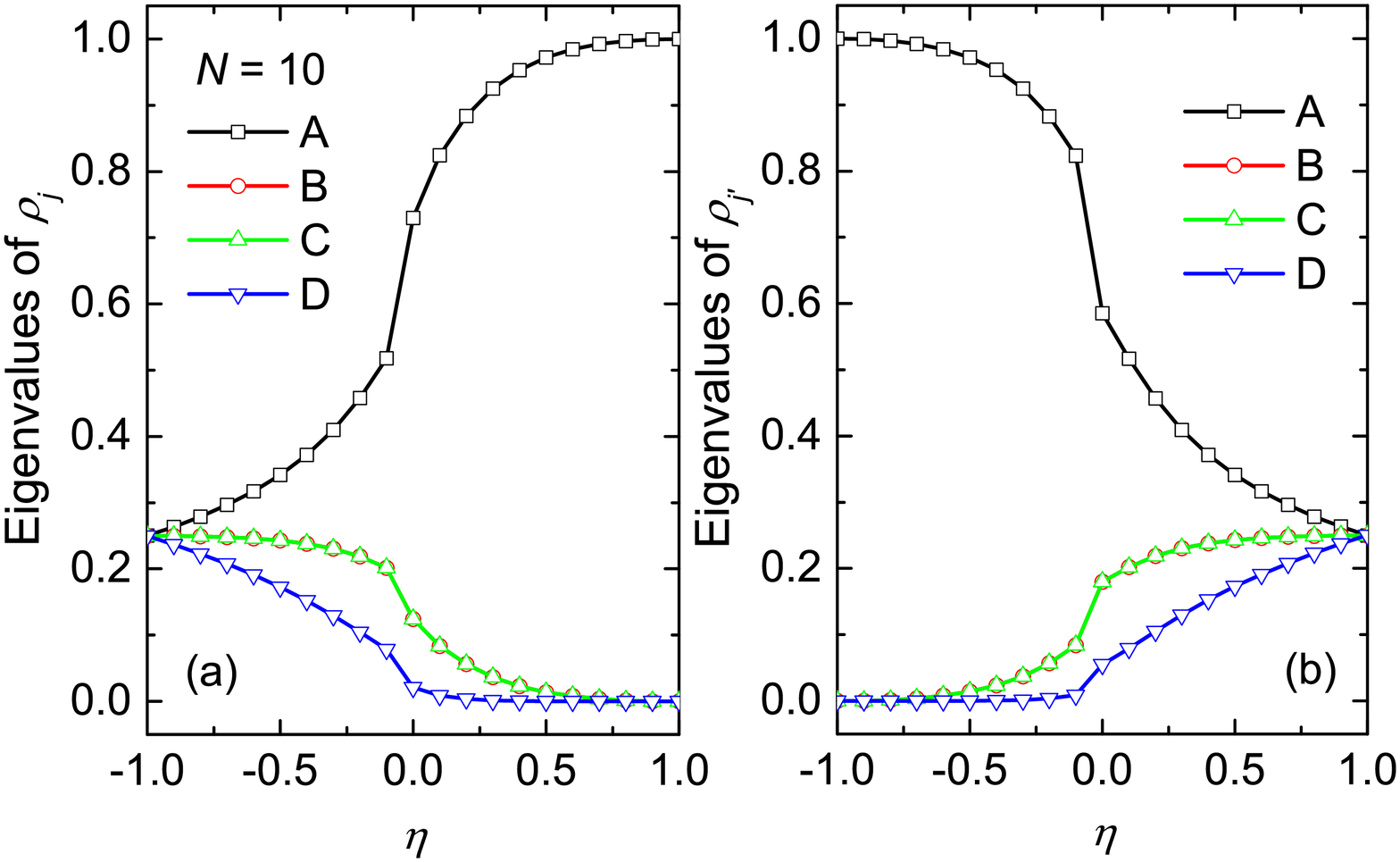}
\caption{\label{fig5} (Color online) Eigenspectrum of the reduced
density matrix calculated with a block consisting of (a) an atom A
and atom B at site $j$, and (b) an atom B at site $j$ and an atom A
at site $j+1$ using PBC for a system of $N=10$. }
\end{figure}

To derive the order parameter, we calculated the single-site reduced density matrix using periodic boundary condition (PBC). In the basis of $\ket{n_{j,A},n_{j,B}}=\{\ket{00},\ket{01},\ket{10},\ket{11}\}$, the reduced density matrix takes the form
\bea
\rho_j=\left(
\begin{array}{cccc}
u & 0 & 0 & 0 \\
0 & v & z & 0 \\
0 & z & v & 0 \\
0 & 0 & 0 & u%
\end{array}%
\right).
\eea
The eigenstates are given by
\bea
|A \rangle &=& \frac{1}{\sqrt{2}} \left( |1 0 \rangle + |0 1 \rangle \right) \nonumber \\
|B \rangle &=& |0 0 \rangle \nonumber \\
|C \rangle &=& |1 1 \rangle  \nonumber \\
|D \rangle &=& \frac{1}{\sqrt{2}} \left( |1 0 \rangle - |0 1 \rangle \right) \nonumber \\
\label{eq:Estates}
& &
\eea
and the corresponding eigenvalues are shown in Fig. \ref{fig5}(a).


For $\eta<0$, the four eigenstates are equally weighted as $\eta \rightarrow -1$. According to our scheme \cite{Gu2013,YU2016}, the order parameter can be defined as
\bea
O_- &=& w_A |A \rangle \langle A | +w_B |B \rangle \langle B| \nonumber\\
&&+ w_C |C \rangle \langle C|+w_D |D \rangle \langle D|.
\eea
Here we have four variables to be fixed but we only have the traceless and cut-off conditions.

\begin{figure}
\includegraphics[width=0.9\columnwidth]{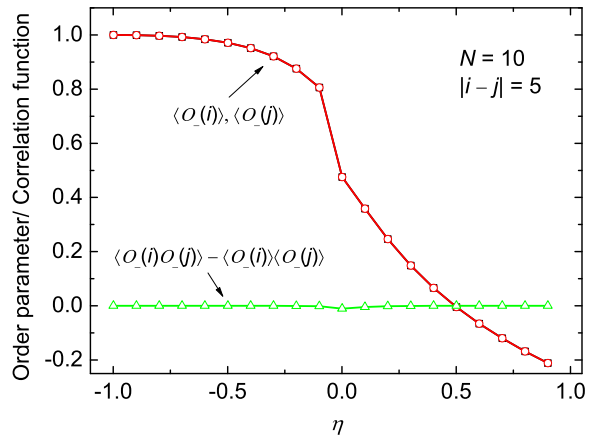}
\caption{\label{fig6} (Color online) Order parameter and the connected correlation function of $O_-$ in Eq. (\ref{eq:On}) as a function of $\eta$.}
\end{figure}

Instead, we may try a different approach by changing the basis used
to define the reduced density matrix. As shown in the previous
section, the Hamiltonian is diagonalized in terms of the $d$ and $f$
fermions at the point $\mu=0,\eta=-1$. At this point the reduced
density matrix is solely contributed by the $|n_f=0,n_d=0 \rangle $
state. The Hamiltonian is trivially diagonal and the eigenvector of
the reduced density matrix is just the eigenvector of the state for
which both $d$ and $f$ are empty. (Unlike in the original
description in terms of the $c_A$ and $c_B$ operators, for which all
four states contribute equally). So the representation of the states
depends on the basis used (meaning which operators we use). Due to
the nature of the topological region, one expects that as long as
the system remains gapped the properties of the system should be
qualitatively the same for all $\eta<0$.

\begin{figure}
\includegraphics[width=0.8\columnwidth]{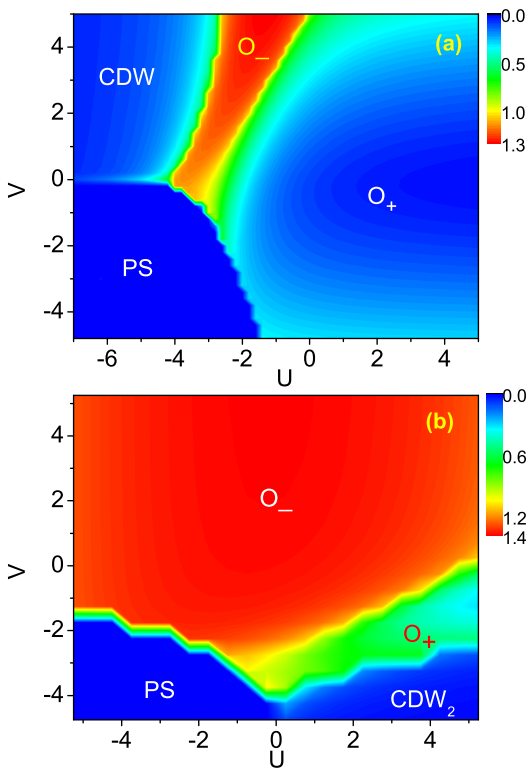}
\caption{\label{fig7} (Color online) Single-site entanglement entropy for
(a)$\eta=0.6$ and (b)$\eta=-0.6$ as a function of $U$ and $V$. In
each region the dominant order parameters are shown. There is a
clear correlation between the entanglement and the order parameter
$O_-$. }
\end{figure}

In the diagonal basis the order parameter is
\bea
O_- &=& |0 0 \rangle \langle 0 0| \nonumber \\
&=& I - |1 0\rangle \langle 1 0| - |0 1\rangle \langle 0 1| - |1 1\rangle \langle 1 1| \nonumber \\
&=& I-f^{\dagger}_j f_j - d^{\dagger}_j d_j -f^{\dagger}_j f_j d^{\dagger}_j d_j. \nonumber \\
& &
\eea
These expressions are local in space. We may now use the relation
between the $d$ and $f$ operators and the original operators in Eq. (\ref{eq:df}). This
is a non-local transformation since it couples site $j$ with the nearest-neighbor
site $j+1$. The operator may now be obtained as
\bea
O_- &=& \frac{3}{2} \left( c_{j+1,A}^{\dagger} c_{j,B} + c_{j,B}^{\dagger} c_{j+1,A} \right)
\nonumber \\
&+& n_{j,B} n_{j+1,A} - \frac{1}{2} \left( n_{j,B} + n_{j+1,A} \right).
\label{eq:On}
\eea

For $\eta>0$, the mutual information is exponentially vanishing and the correlation is not captured by considering the single-site block with atoms A and B. However, one could take the block consisting of an atom B at site $j$ and an atom A at site $j+1$. The mutual information obtained would be the mirror image of that in Fig. \ref{fig4} about $\eta=0$. The eigenspectrum in this case is shown in Fig. \ref{fig5}(b). Carrying out similar analysis as above, the order parameter takes the form of Eq. (\ref{eq:On}), but with the index $\{j+1,A\}$ and $\{j,B\}$ being replaced by $\{j,B\}$ and $\{j,A\}$, respectively. We have
\bea
O_+ &=& \frac{3}{2} \left( c_{j,B}^{\dagger} c_{j,A} + c_{j,A}^{\dagger} c_{j,B} \right)
\nonumber \\
&+& n_{j,A} n_{j,B} - \frac{1}{2} \left( n_{j,A} + n_{j,B} \right).
\label{eq:Op}
\eea

In Fig. \ref{fig6}, we show the results for the order parameter $O_-$ and its correlation function as a function of the dimerization $\eta$ (for $O_+$ could be obtained by taking the mirror image about $\eta=0$). By construction we see that the order parameter is dominant in its intended region of applicability and change continuously
from a finite value towards zero or small values as we move to the opposite region.
However, since the system is actually not ordered the connected correlation function $\langle O_iO_j\rangle-\langle O_i\rangle\langle O_j\rangle$
vanishes in all regimes.

\subsection{Discussion}
Regarding the above derivations, note that the dominating eigenstate of the
reduced density matrix is given by a single state in the basis chosen. This is in contrast
with the case of a continuous phase transition in which symmetric eigenstates
would be resulted in a finite system. Consequently, we did not apply the
traceless condition and the cut-off conditions in the derivation. The resulted
order parameters, despite showing a sharp change around the quantum critical point, do not behave as conventional ones (finite in the "ordered" phase and goes to zero in the "disordered" phase).
The order parameters are defined accross two lattice locations: at the same site between
the two types of (sublattice) locations $A$ and $B$ for $\eta>0$, and linking two locations
$A$ and $B$ between neighboring sites. A vanishing order parameter may be constructed
summing two consecutive links with opposite signs as in the bond-order (BOW) parameter
\cite{benthien}.

\begin{figure*}
\includegraphics[width=0.9\textwidth]{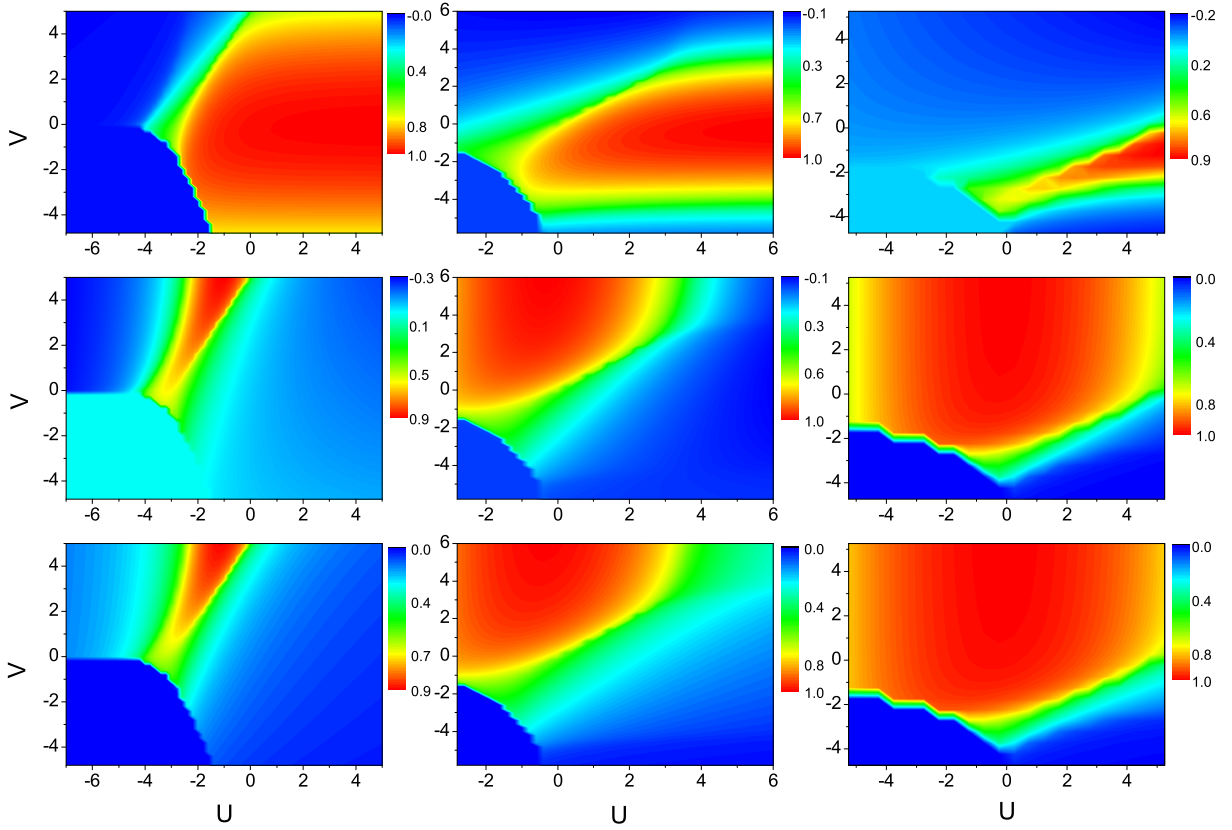}
\caption{\label{fig8} (Color online) Order parameters suitable for $\eta=0.6, \eta=0, \eta=-0.6$ (from
left to right). In the first row for $O_{+}$ given in Eq.
(\ref{eq:Op}). In the second row $O_{-}$ given in Eq. (\ref{eq:On}).
In the third row for $m_{\textrm{BOW}}$ given in Eq. (\ref{mbow}). }
\end{figure*}


The order parameter in the topological region is similar to the BOW
order parameter, with a few more terms related to the densities at
sites $A$ and $B$.

\section{\label{sec:level5} Effect of interactions}

Adding interactions is interesting because i) allows a generalization of the procedure of
finding order parameters to a problem that is now interacting
and to determine how the
interactions affect the choice of order parameter(s) to describe the
various phases,
and ii) may change the
topological properties determined for the non-interacting
system.

\subsection{DMRG results for order parameters}

We add a local Hubbard-$U$ like term (coupling two electrons at the same
site but in two different sublattices, $A$ and $B$) and/or a $V$-term coupling
two electrons at nearest-neighbor sites.

In the presence of interactions the model Hamiltonian is chosen as
\bea
H &=& -\sum_j \left[ \left(1+\eta \right) c_{j,A}^{\dagger} c_{j,B} + \left(1-\eta \right) c_{j,B}^{\dagger} c_{j+1,A}
+ h.c. \right] \nonumber \\
&+& U \sum_j n_{j,A} n_{j,B} + V \sum_j n_{j,B} n_{j+1,A}
\label{eq:HI} \eea We calculate, using the infinite density matrix
renormalization group method \cite{white} with PBC, the entanglement entropy and various order parameters as
a function of $\eta,U$ and $V$. The truncation error is set to
less than $10^{-7}$ in our calculations. The system size simulated is $N=86$ unless otherwise specified.
Specifically we calculate
in addition to the order parameters $O_+$ and $O_-$, a bond-order
parameter defined on a link
\be
m_{\textrm{BOW}} = \langle \left(
c_{j+1,A}^{\dagger} c_{j,B} + H.c. \right)\rangle. \label{mbow}
\ee

\begin{figure}
\includegraphics[width=0.95\columnwidth]{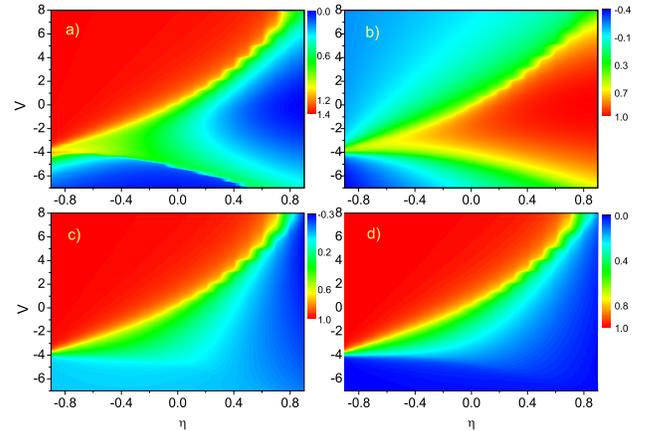}
\caption{\label{fig9} (Color online) (a) Single-site entanglement entropy, and order parameter (b) $O_{+}$, (c) $O_{-}$,
(d) $m_{\textrm{BOW}}$ for $U=0$ as
a function of $V$ and $\eta$.}
\end{figure}

We begin by accessing the effect of the interaction on the
single-site entanglement entropy shown in Fig. \ref{fig7} for a
point in the trivial region with $\eta=0.6$ and another point in the
topological region with $\eta=-0.6$ as a function of the
interactions $U$ and $V$. The smoothness of the phase boundaries were limited by the point density of the driving parameters.
As shown in Fig. \ref{fig3}, the entanglement
entropy is large in the topological region with no interactions. The
presence of repulsive inter-site interaction $V$ does not change the entanglement entropy. However, if $V<0$, the entanglement entropy is reduced, particularly when $U<0$. The decrease of the entanglement entropy is more gradual if $U>0$. In the trivial region ($\eta>0$), the entanglement is also large in the regime
where the order parameter $O_-$ has a large value.

In Fig. \ref{fig8} we compare various order parameters for three points: one in the topological
trivial region ($\eta=0.6$), one at the transition point where the dimerization $\eta=0$, and one in the topological region ($\eta=-0.6$).
The results are presented as a function of
the interactions $U$ and $V$.

A first comment is that there is some interpolation between the
topological and the trivial regions as one crosses the transition
point. At least from the point of view of the order parameters,
there does not seem to be a clear distinction between the two
topologically different regimes. This is consistent with the idea
that a topological transition is subtle and is not straightforwardly
associated with a change of some order parameters. However, it is the
purpose of the choice of order parameters by analysing the reduced
density matrix eigenstates and eigenvlaues to construct order
parameters without the necessary use of any symmetry breaking
arguments. The similarity of the order parameters $m_{\textrm{BOW}}$ and
$O_{-}$ might lead to the expectation that, at least in this case,
the method is actually capturing the traditional types of order (as
also revealed in the mutual information results) instead of some
form of topological property.

The order parameters $m_{\textrm{BOW}}$ and $O_{-}$ are particularly expanded in the phase
diagram in the topological regime. Their extension decrease as one crosses over to
the trivial region, as might be expected since they are particularly suited to the
topological region.

As one crosses to the trivial region, the effect of $U$ (local term) becomes
more prominent as evidenced by the local nature of the order parameter $O_{+}$.
As expected the effect of the interactions is smaller in the trivial regime where
extended regions in the phase diagram result in a large value of this order parameter.

Given that the effect of the local interaction, $U$, is small particularly in the
topological regime, we take $U=0$ in Fig. \ref{fig9} and study the effect on the order
parameters and the single-site entanglement entropy as a function of $V$ and $\eta$.
We clearly see the dominance of the order parameters $O_{-}$ and $m_{\textrm{BOW}}$ in the topological
region of $\eta<0$ and of the order parameter $O_{+}$ in the trivial regime $\eta>0$.

The order parameters considered so far are only defined in single links.
In order to probe possible long-range order we need to consider two cells
containing at least two consecutive links. One may consider
possible related order parameters defined as follows:
\begin{align}\label{eq:bow}
O_{\textrm{BOW}}=\left\langle
\left(c^\dagger_{j+1,B}c_{j,A}+H.c.\right)-\left(c^\dagger_{j,B}c_{j,A}+H.c.\right)\right\rangle
\end{align}
for bond-ordering (BOW) and
\begin{align}\label{eq:cdw}
O_{\textrm{CDW}}=\frac{1}{2}\left\langle
\left(n_{j+1,A}+n_{j+1,B}-1\right)-\left(n_{j,A}+n_{j,B}-1\right)\right\rangle
\end{align}
for charge-density wave (CDW) ordering.

\begin{figure}
\includegraphics[width=1.0\columnwidth]{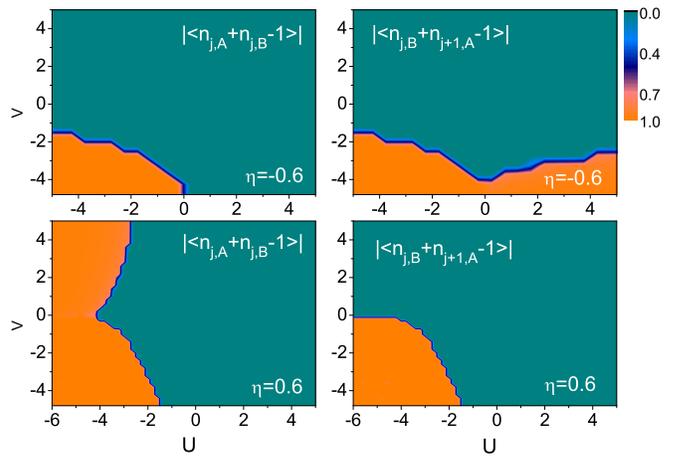}

\caption{\label{fig10} (Color online) Electron number density analysis on different links: (left column) A and B at the same site, (right column) A and B linked by different sites. The fist row is for $\eta=-0.6$ and the second row is for $\eta=0.6$. }
\end{figure}

\begin{figure}
\includegraphics[width=0.85\columnwidth]{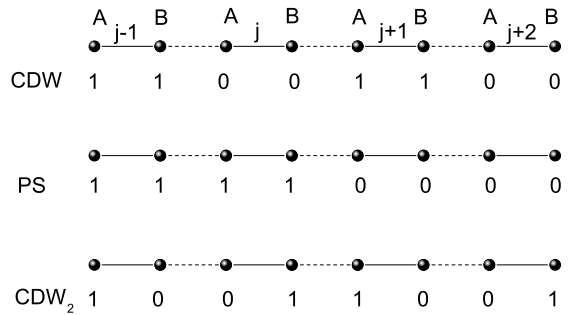}
\caption{\label{fig11} Possible electron
configuration for the three quantum states indicated in Fig.
\ref{fig10}. }
\end{figure}

The analysis of several partial links is shown in Fig.~\ref{fig10}, which leads to the conclusion that the electron configurations may be described by the scenario show in Fig. \ref{fig11}.

\begin{figure}
\includegraphics[width=0.8\columnwidth]{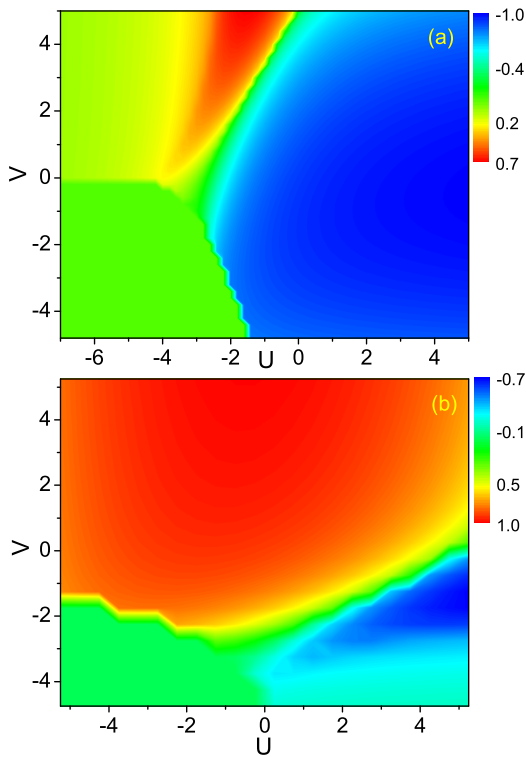}
\caption{\label{fig12} (Color online) BOW order parameter of eq.
(\ref{eq:bow}) for (a) $\eta=0.6$ and (b) $\eta=-0.6$. }
\end{figure}

\begin{figure}
\includegraphics[width=1.0\columnwidth]{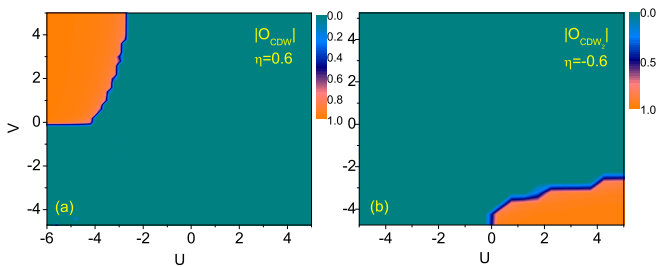}
\caption{\label{fig13} (Color online)(a) CDW order parameter of eq.
(\ref{eq:cdw}) for $\eta=0.6$. Because $O_{CDW}=0$ at the whole
parameter region for $\eta=-0.6$, we do not show it here. (b)
$\mathrm{CDW}_2$ order parameter of eq.~(\ref{eq:cdw2}) for
$\eta=-0.6$. }
\end{figure}

The order parameters related to BOW and CDW as a function of $U$ and $V$ are presented in Fig. \ref{fig12} and Fig. \ref{fig13} respectively. The results for the BOW order
parameter are consistent with those obtained from the derived order parameters $O_+$ and $O_-$ in the previous section. In the case without interactions ($U=0,V=0$), $O_+$ dominates, and $O_{\textrm{BOW}}<0$ and is close to $-1$ in the trivial region. While in the topological region, the order parameter $O_-$ dominates, and $O_{\textrm{BOW}}>0$ and is close to $1$. The effect of interactions is similar to the one observed for the link order parameters. On the other hand, the result in \ref{fig13}(a) shows that the inter-site repulsion ($V>0$) and the intra-site attraction ($U<0$) between the electrons favor the CDW order in Fig. \ref{fig11}. According to the analysis on electron configurations, we could define another order parameter
\begin{eqnarray}
\label{eq:cdw2}
O_{\textrm{CDW$_2$}}&=&\frac{1}{2}\left\langle
\left(n_{j+1,B}+n_{j+2,A}-1\right)\right.\nonumber\\
&&\hspace{18pt}\left.-\left(n_{j,B}+n_{j+1,A}-1\right)\right\rangle,
\end{eqnarray}
which describes the charge-density wave of the links between adjacent sites. The nonzero values appear at the $\eta=-0.6$ case as shown in Fig.~\ref{fig13}(b). We therefore conclude the nonzero region, which corresponds to the $\mathrm{CDW}_2$ region in Fig.~\ref{fig7}, belongs to the $\mathrm{CDW}_2$ order.

\subsection{Berry phase in the presence of interactions}

To see if and how the topology changes one needs to look at edge states
(using open boundary conditions) or looking at topological invariants
(using periodic boundary conditions).
One method to calculate a topological invariant involves
calculating the Green's function and using the definitions of the invariants \cite{manmana}.
Another possibility to study a topological invariant
is to calculate the Berry phase \cite{berry1,berry2}.

\begin{figure}
\includegraphics[width=1\columnwidth]{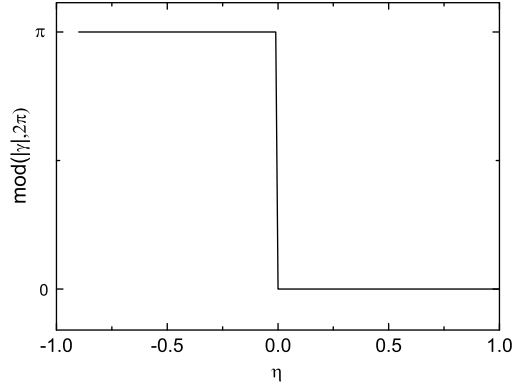}
\caption{\label{fig14} Berry phase as a function of
$\eta$ for the non-interacting case. }
\end{figure}

Using twisted boundary conditions we can calculate the Berry phase
which is a topological invariant that reveals the topological nature
of the system. Imposing a phase of $\phi$ in the boundary conditions
the Berry phase may be calculated as
\be
\gamma = -i \int_0^{2\pi}
\langle \psi(\phi) | \frac{\partial}{\partial \phi} \psi(\phi)
\rangle.
\ee
In order to calculate the Berry phase, it is more
convenient to discretize the range of phase values into $M$ points, i.e. $\phi_1,\phi_2,\cdots,\phi_M$.
Defining the link variable $U(\phi_l) = \psi^*(\phi_l)
\psi(\phi_{l+1})$ and summing over $\phi_l$, we may obtain the Berry
phase as \be \gamma = -i \sum_{l=1}^M \ln U(\phi_l). \ee Consider
first the non-interacting case. In Fig. \ref{fig14} the results for
the Berry phase as a function of $\eta$ in the absence of
interactions is shown. In the topological region the Berry phase is
$\pi$ and in the trivial phase it vanishes, as expected of the
topological transition discussed above. We performed ED for small systems to calculate the overlap
between the groundstate at nearby values of the phase imposed by
twisted boundary conditions.

\begin{figure*}
\includegraphics[width=0.45\textwidth]{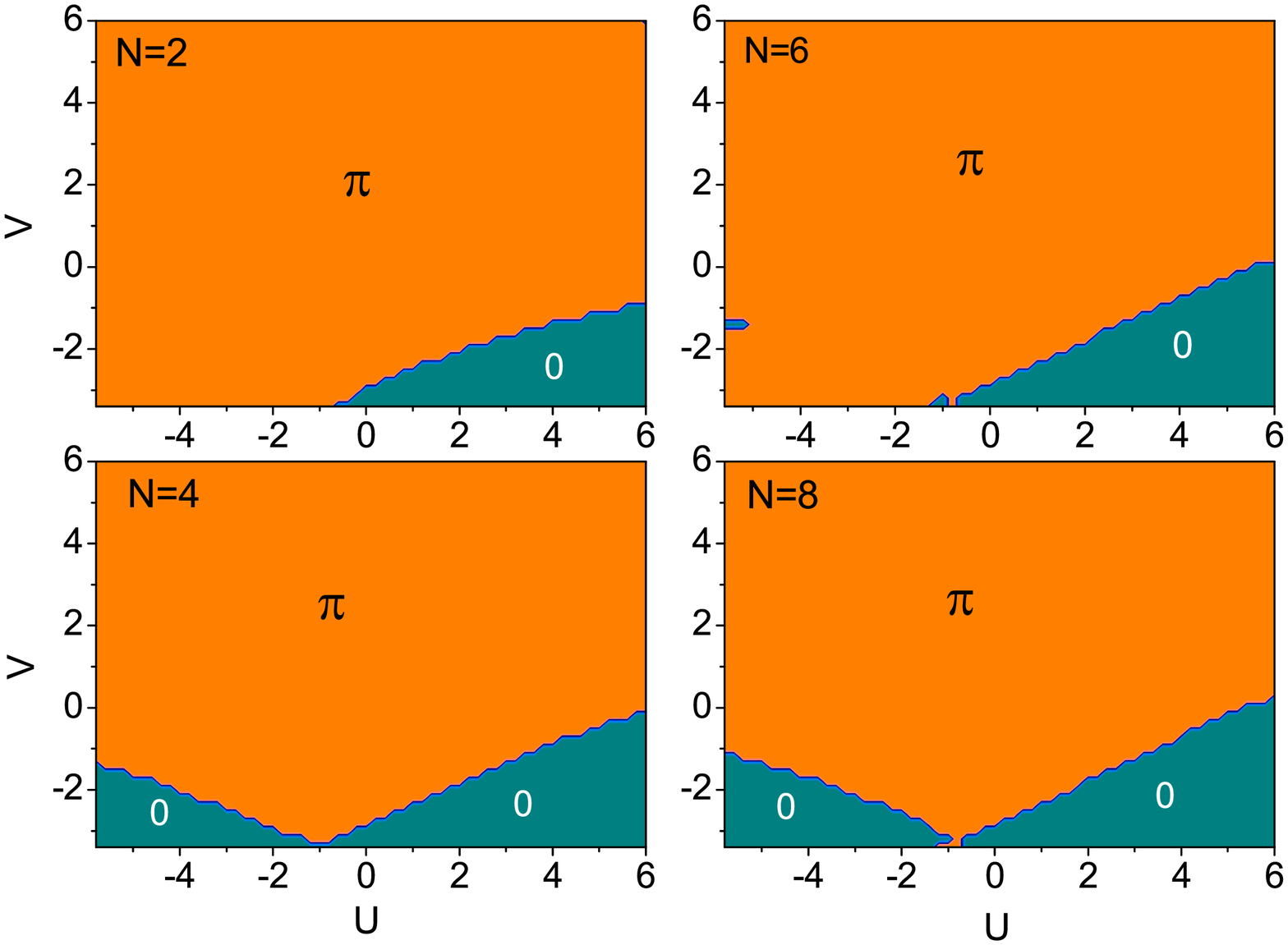}
\includegraphics[width=0.45\textwidth]{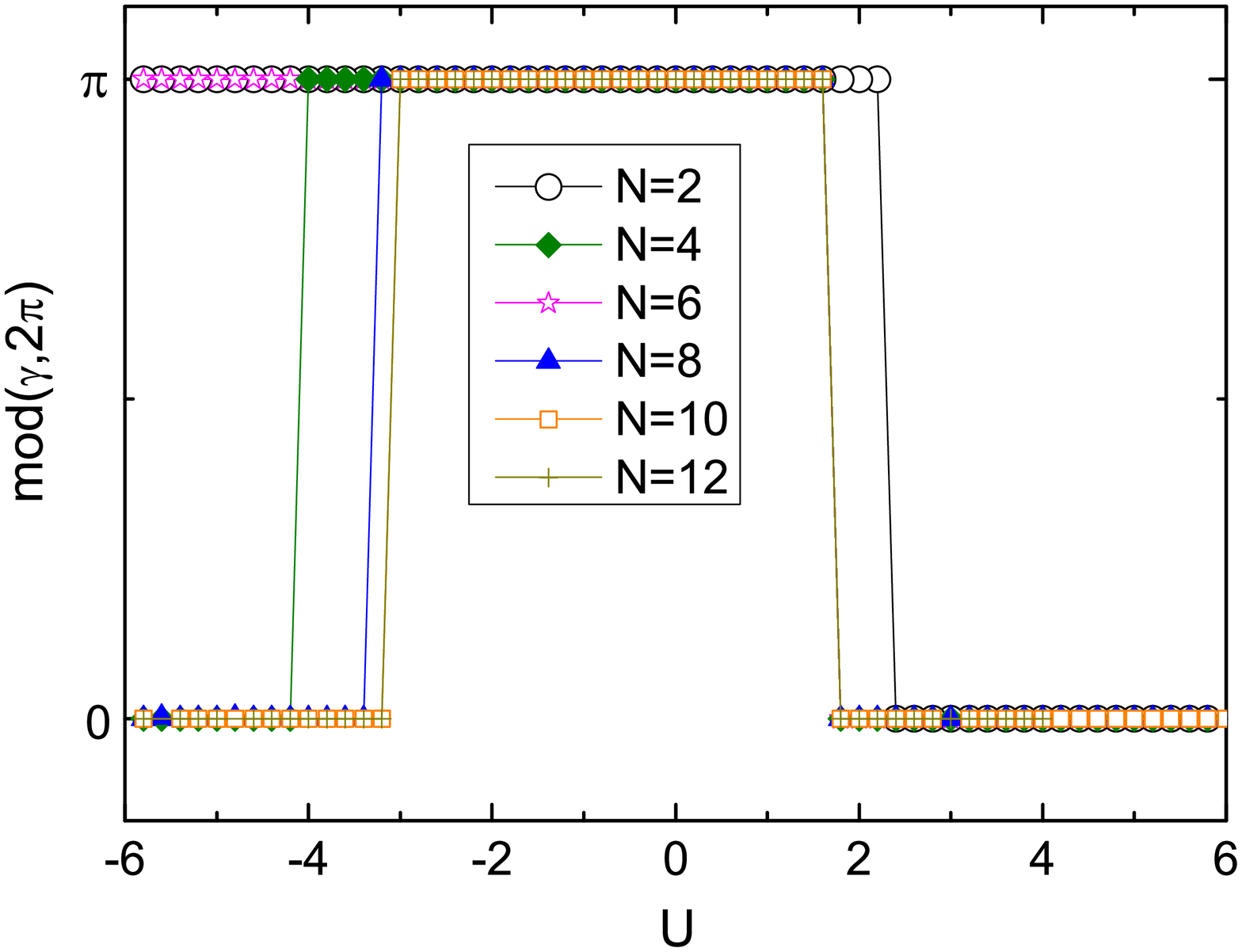}
\caption{(Color online)
Berry phase as a function of the interactions for $\eta=-0.6$ for
various system sizes.
\label{fig15}
}
\end{figure*}

In Fig. \ref{fig15} we analyze the size dependence of the Berry
phase for $\eta=-0.6$. As is well known, the $V$ term may induce
bond-order which is a characteristic feature of the topological
region. Starting from the topological region we see that both
negative $V$ and negative $U$ affect the topology and a trivial
regime characterized by the vanishing Berry phase may appear as a
result. In Fig. \ref{fig15} we also consider the size dependence of
the results for $\eta=-0.6$. Due to finite size effects there is a
$4N$, $4N+2$ alternancy. In the very large size limit the results
converge to the $4N$-case, as shown in the right plot of
Fig.~\ref{fig15}, where the curves of Berry phase for $V=-2.0$ tends
to the same value as $N$ is larger than 10 (system sizes up to $N=12$ is
considered here). 

\begin{figure}
\includegraphics[width=0.8\columnwidth]{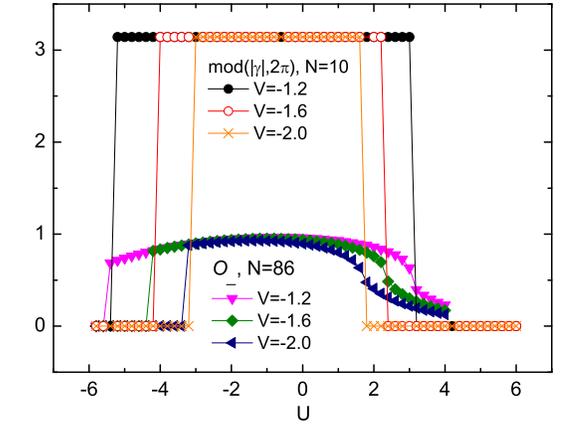}
\caption{\label{fig16} (Color online) Berry phase and order
parameter $O_{-}$ as a function of $U$ under different $V$. The
dominant region of $O_{-}$ coincides with the topological phase
indicated by the $\pi$ value of Berry phase.}
\end{figure}

To illustrate the relationship between the topological phase and the
derived order parameter $O_{-}$, Fig.~\ref{fig16} plots the Berry
phase and $O_{-}$ under the same parameter's conditions. The points where $O_-$ changes dramatically is consistent with the edge of the
topological $\pi$ region of the Berry phase. Therefore, the dominant
region of $O_{-}$ indeed describes the topologically non-trivial
phase. In addition, the Berry phase for $\eta=0.6$ is also $\pi$ when
$O_-$ is dominant due to the effect of interactions (negative U
positive V, as shown in Fig.~\ref{fig12}). This confirms the
appearance of topology due to the interactions having started at
$U=0$ and $V=0$ from a trivial phase. Also, it extends the
relationship between a Berry phase of $\pi$ and a non-zero $O_-$.

\subsection{Reduced density matrix and order parameters in the presence of interactions}

\begin{figure}
\includegraphics[width=1.1\columnwidth]{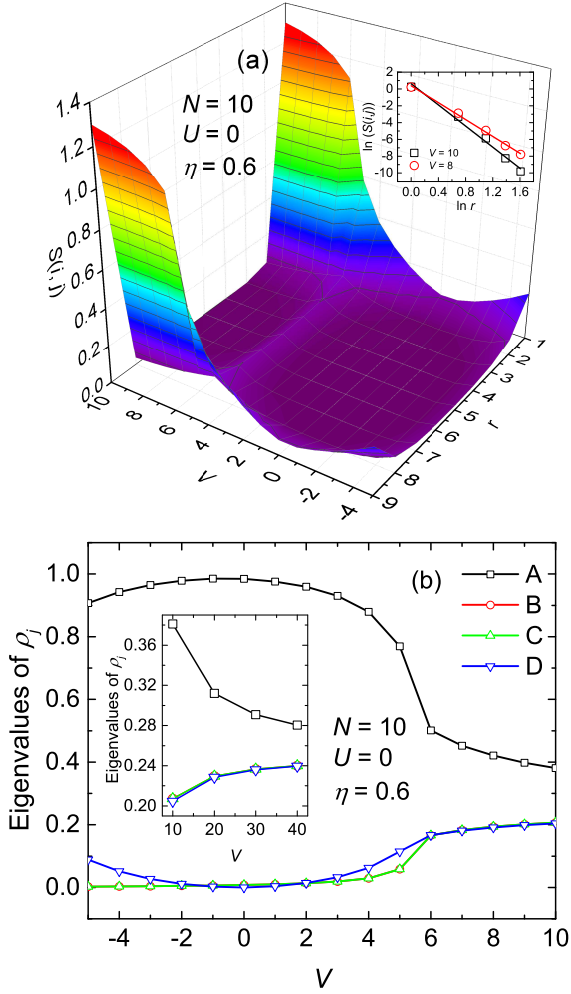}
\caption{\label{fig17} (Color online) (a) Mutual information as a
function of $V$ and distance $r=|i-j|$ for $U=0,\eta=0.6$. Inset shows that the mutual information decays algebraically with the distance. (b) The eigenspectrum of the reduced density matrix $\rho_j$ as a function of $V$. Inset shows the asymptotic weight of the eigenstates of the reduced density matrix in the large $V$ limit.}
\end{figure}

\subsubsection{$\eta=0.6$}

\begin{figure}
\includegraphics[width=1.1\columnwidth]{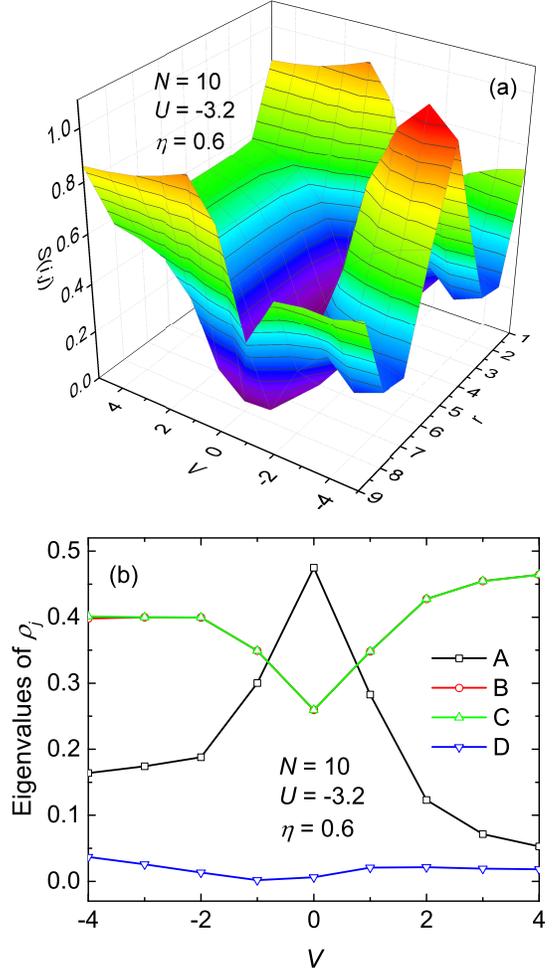}
\caption{\label{fig18} (Color online)(a) Mutual information as a
function of $V$ and distance $r$ for $U=-3.2,\eta=0.6$. (b) The eigenspectrum of the reduced density matrix $\rho_j$ as a function of $V$.} \end{figure}

Let us first consider varying $V$ along the path of fixed $U=0$. The
mutual information as a function of $V$ and the distance $r$
calculated for $U=0$ with PBC is shown in Fig. \ref{fig17}(a). As
indicated by the log-log plot in the inset, the mutual information
decays algebraically with the distance and we could argue that there
exist a long-range correlation in the system for $V>6$.

Figure \ref{fig17}(b) shows the eigenvalues of the states in
Eq.~(\ref{eq:Estates}) of the single-site reduced density matrix.
For $V>6$, the contribution of the four eigenstates are similar and
they are almost equally weighted in the large $V$ limit (inset of
\ref{fig17}(b)). This is the same as the case for $\eta<0$ in the
non-interacting system. Following the same argument in Sec.
\ref{sec:4C}, the order parameter for this phase is $O_{-}$ in Eq.
(\ref{eq:On}).

For $V<6$, from the result of single-site entanglement show in
Fig.~\ref{fig7}, the system is in the same phase as that of $U=V=0$.
The behavior of the reduced density matrix eigenspectrum is the same
as for the non-interacting case for $\eta>0$. The order parameter in
this regime is $O_{+}$ as obtained in Eq. (\ref{eq:Op}).

\begin{figure}
\centering
\includegraphics[width=1.05\columnwidth]{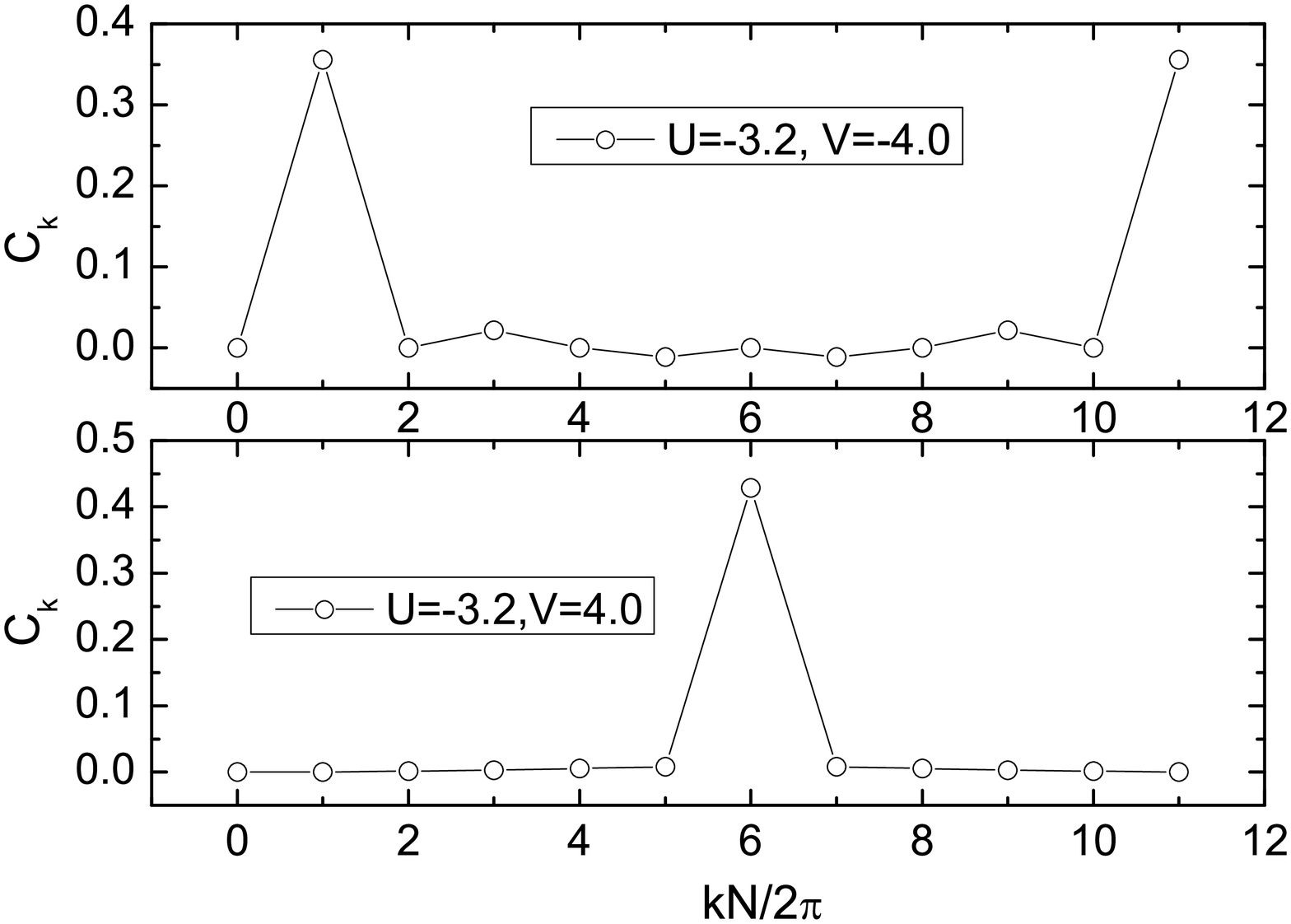}
\caption{\label{fig19} The correlation function of $O_3$ in the momentum space for $U=-3.2$ and $V=\pm4$. Here $k=2m\pi/N$, where $m=0,1, \cdots,N-1$.}
\end{figure}

To analyze the different possible phase regions in Fig.~\ref{fig7},
we next consider the path along fixed $U=-3.2$. Figure \ref{fig18}
shows the mutual information as a function of $r$ and $V$.
Obviously, the mutual information for $V>2$ and $V<-2$ (corresponding to CDW and PS phases in Fig.~\ref{fig7} respectively) is non-vanishing at a long distance. In Fig. \ref{fig18}(b), the eigenspectrum of the reduced density
matrix is dominated by the states $\ket{B}$ and $\ket{C}$ for both regions. For large enough system, the weight of state
$\ket{A}$ would be suppressed to zero (our DMRG results, which we do
not show here, indicate this). We can define the order parameter as
\bea O_3=w_B\ket{B}\bra{B}+w_C\ket{C}\bra{C}. \eea Using the
traceless condition, we have $w_B=-w_C$ and applying the cut-off
condition, i.e. $w_B=1$, we have \bea
O_3&=&\ket{00}\bra{00}-\ket{11}\bra{11}, \nonumber\\
&=& 1- (n_{j,A}+n_{j,B}), \label{eq:O3} \eea which is indeed the
order parameter for the CDW and PS phases. To further distinguish
the two phases, we can consider the correlation function in the
momentum space shown in Fig. \ref{fig19}. The correlation function
peaks at $2\pi/N$ (and $2\pi(N-1)/N$ as a result of PBC) and $\pi$
for $V<-2$ and $V>2$, respectively. It indicates that the electronic
configuration has a wavelength of half of the lattice in the former
case and of two sites in the latter case. This is consistent with
our deduction for the PS and CDW phases illustrated in
Fig.~\ref{fig11}. In addition, the behavior of the mutual
information in Fig.~\ref{fig18}(a) also reflects the difference of
the two phases. In the PS phase, the largest correlation appears
between two local sites separated by half of the lattice.

\subsubsection{$\eta=-0.6$}

Consider now the case $U=4$. Figure \ref{fig20} shows the mutual information as a function of $V$ and $r$. In the case of $V>0$, a correlation emerges between
the end points of the system. Note from Fig. \ref{fig8} that in this
regime $O_-$ also becomes large. On the other hand, for $V<0$, the Berry phase
results of Fig. \ref{fig15} do not indicate a topological phase,
consistent with the separation of the two types of properties. For
negative $V<-2.5$ the correlation extends all over the system, as
expected.

Figure \ref{fig21}(a) shows the eigenvalues of the states $\{A, B, C, D\}$ of the reduced density matrix. For $V>0$, all the eigenstates have non-negligible weight. Let us once again transform into the basis defined by the $d$ and $f$ operators using Eq. (\ref{eq:cdf}). Under PBC and keeping the number of electrons equal to the number of sites (half-filling) the
transformed Hamiltonian reads $H=H_1+H_2$ where
\bea
H_1 &=& \frac{t}{2} \left(1+\eta \right) \sum_{j=1}^N \left(
f_j^{\dagger} f_{j+1}^{\dagger} +f_{j+1} f_j + f_j^{\dagger} f_{j+1} +f_{j+1}^{\dagger} f_j \right)
\nonumber \\
 &+& \frac{t}{2} \left(1+\eta \right) \sum_{j=1}^N \left(
d_j^{\dagger} d_{j+1}^{\dagger} +d_{j+1} d_j + d_j^{\dagger} d_{j+1} +d_{j+1}^{\dagger} d_j \right)
\nonumber \\
&+& \left[ t(1-\eta) +\frac{V}{2} \right] \sum_{j=1}^N \left( n_j^f + n_j^d \right)
\nonumber \\
&-& V \sum_{j=1}^N n_j^f n_j^d
+ N \left[ \frac{U}{4} -t(1-\eta) \right],
\eea
and
\bea
H_2 &=&
-\frac{U}{4} \sum_{j=1}^N \left(
f_{j}^{\dagger} f_{j+1}d_{j}^{\dagger} d_{j+1}
+f_{j+1}^{\dagger} f_{j}d_{j+1}^{\dagger} d_{j}
\right.
\nonumber \\
&+& \left.
f_{j+1}^{\dagger} f_{j}d_{j}^{\dagger} d_{j+1}
+f_{j}^{\dagger} f_{j+1}d_{j+1}^{\dagger} d_{j}
\right.
\nonumber \\
&+& \left.
f_{j} f_{j+1}d_{j+1}^{\dagger} d_{j}^{\dagger}
+f_{j+1}^{\dagger} f_{j}^{\dagger}d_{j} d_{j+1}
\right.
\nonumber \\
&+& \left.
f_{j}^{\dagger} f_{j+1}^{\dagger}d_{j}^{\dagger} d_{j+1}^{\dagger}
+f_{j+1} f_{j} d_{j+1} d_{j}
\right.
\nonumber \\
&-& \left.
f_{j+1}^{\dagger} f_{j}^{\dagger}d_{j}^{\dagger} d_{j+1}
-f_{j} f_{j+1} d_{j+1}^{\dagger} d_{j}
\right.
\nonumber \\
&-& \left.
f_{j}^{\dagger} f_{j+1} d_{j+1}^{\dagger} d_{j}^{\dagger}
-f_{j+1}^{\dagger} f_{j} d_{j} d_{j+1}
\right.
\nonumber \\
&-& \left.
f_{j+1}^{\dagger} f_{j} d_{j+1}^{\dagger} d_{j}^{\dagger}
-f_{j}^{\dagger} f_{j+1} d_{j} d_{j+1}
\right.
\nonumber \\
&-& \left.
f_{j} f_{j+1} d_{j}^{\dagger} d_{j+1}
-f_{j+1}^{\dagger} f_{j}^{\dagger} d_{j+1}^{\dagger} d_{j}
\right).
\eea

In the basis of $|n_j^f,n_j^d\rangle$ the eigenstates of the single-site reduced density matrix take the form
\bea
|A^{\prime} \rangle &=& \alpha |10\rangle + \beta |01\rangle, \nonumber \\
|B^{\prime} \rangle &=& |00\rangle,
\nonumber \\
|C^{\prime} \rangle &=& |11\rangle,
\nonumber \\
|D^{\prime} \rangle &=& \beta |10\rangle - \alpha |01\rangle. \eea
As shown in Fig. \ref{fig21}(b), the state $\ket{B'}$ is dominant in the region $V>0$ and we would arrive at the same order parameter, i.e. $O_-$, as in the case of no interaction. The order parameter in this topological region prevails and is not affected by the interactions.

For $V<-2.5$, states $\ket{A}$ and $\ket{D}$ dominant and let us define the order parameter as
\bea
O_{+,1} &=& w_A |A\rangle \langle A | +w_D |D\rangle \langle D| \nonumber \\
&=& \frac{1}{2} \left(w_A + w_D \right) \left( |10\rangle \langle 10| + |01\rangle \langle 01| \right)
\nonumber \\
&+& \frac{1}{2} \left(w_A - w_D \right) \left( |10\rangle \langle 01|+
|01 \rangle \langle 10| \right).
\eea
Without the loss of generality, assume that the weights of the two states $\ket{A}$ and $\ket{D}$ tends to an asymptotic value of 0.5 with some probably chosen values of $U$ and $V$ within the same phase.  Traceless condition then gives $w_A=-w_D$ and setting $w_A=1$, we have
\bea
O_{+,1} = c_{j,A}^{\dagger} c_{j,B} + c_{j,B}^{\dagger} c_{j,A}.
\eea

A remark here is that if one consider the form of Hamiltonian in Eq.
(\ref{eq:HI}), the case of a positive $\eta$ is equivalent to the
case of a negative $\eta$ with the role of $U$ and $V$ interchanged
and $\{j,A\},\{j,B\}$ being replaced by $\{j,B\},\{j+1,A\}$
respectively. Therefore, for the case $V<-2.5$, one can also take
$O_3$ in Eq. (\ref{eq:O3}) but with the index replaced as the order
parameter. That gives \bea O_{+,2}= 1- (n_{j,B}+n_{j+1,A}),
\label{eq:Op2} \eea which is consistent with the $\mathrm{CDW}_2$
order parameter. In addition, the linear combination of $O_{+,1}$
and $O_{+,2}$ can also be an order parameter.

For $-2.5<V<0.5$, note that the eigenspectrum is similar to the one show in the $V<6$ regime in Fig. \ref{fig17}(b) and supplemented with the above argument, the order parameter for this phase is given by $O_{+}$.

\begin{figure}[tbp]
\centering
\includegraphics[width=1.0\columnwidth]{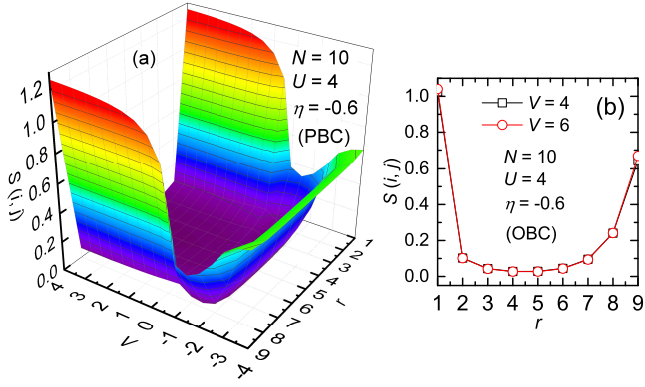}
\caption{{\label{fig20} (Color online) (a) Mutual information as a
function of $V$ and distance $r$ for $U=4,\eta=-0.6$ calculated with PBC. (b) Mutual information as a function of $r$ for $V=6,8$ calculated with OBC. }
\label{fig20}}
\end{figure}

\begin{figure}
\includegraphics[width=1\columnwidth]{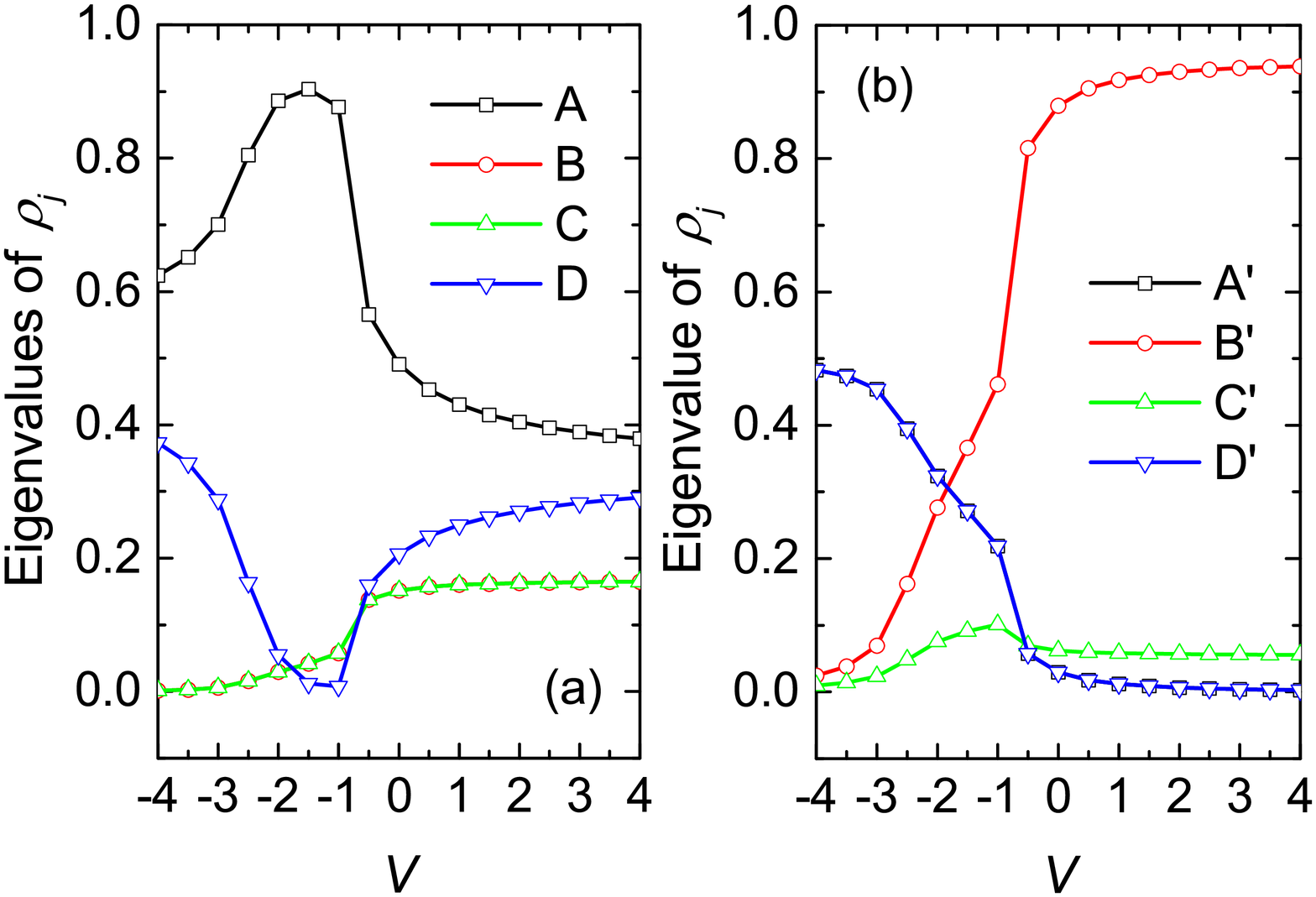}
\caption{{\label{fig21} (Color online) Eigenspectrum of the single-site reduced density matrix in the (a) original basis, and (b) rotated basis as a function of $V$. Here $N=10$, $U=4$ and $\eta=-0.6$.}
\label{fig21}}
\end{figure}

\section{\label{sec:level6} Conclusions}
Using the method proposed from the reduced density matrix\cite{Gu2013}, we
construct the potential order parameters for a condensed matter system,
especially for the topologically non-trivial phase, which can not be
described by those general order parameters derived from the
landau's symmetry breaking theory. We first study the two-band
spinless fermions SSH model. In this simple model, a topological
phase transition exists. We calculate the entanglement entropy, which
clearly identify the quantum critical point. Analyzing the mutual
information and one-site reduced density matrix, we get a local
order parameter for the trivial phase. Furthermore, through an
appropriate change of basis by representing the Hamiltonian in
a Majorana fermion basis, we reduced the number of eigenvalues that
contribute significantly and construct the non-local order parameter
$O_{-}$ for the topologically non-trivial phase.

We then consider the case when the interactions $U$ and $V$ are
added. The entanglement entropy results capture a rich ground state phase diagram
on the $U-V$ plane. Through analysing the electron
configurations, we identify the $\mathrm{PS}$, $\mathrm{CDW}$, and
$\mathrm{CDW}_2$ phases, and give the order parameter for the
$\mathrm{CDW}_2$ state. The order parameters for various quantum
phases are deduced according to the method of analyzing the mutual
information and the reduced density matrix spectra. In addition,
comparing with the dominant regions of different order parameters,
we conclude that the topologically trivial and non-trivial quantum
phases described by $O_{+}$ and $O_{-}$, respectively, could exist
in a wide range of parameter space. Moreover, the topology of the
system affected by the interactions is verified by the Berry phase
results, and the effectiveness of the deduced order parameter
$O_{-}$ in describing the topological quantum phase is further proved.

\begin{acknowledgments}
We acknowledge support from NSAF U1530401, National Natural Science
Foundation of China under Grant No. 11104009, President
Foundation of University of Chinese Academy of Sciences under Grant
No. Y35102DN00, partial support from FCT (Portugal) through Grant
UID/CTM/04540/2013, and computational resources from the Beijing Computational Science Research Center.
\end{acknowledgments}


\begin{thebibliography}{}

\bibitem{Sachdev2000}
S. Sachdev, {\it Quantum Phase Transitions}, (Cambridge University Press,
Cambridge, UK, 2000).

\bibitem{Carr2011}
L. Carr, {\it Understanding Quantum Phase Transitions}, (CRC Press, 2011).

\bibitem{Wen2004}
X. G. Wen, {\it Qunatum Field Theory of Many-body Systems}, (Oxford
University, New York, 2004).

\bibitem{Amico2008}
L. Amico, R. Fazio, A. Osterloh, and V. Vedral, Rev. Mod. Phys. \textbf{80}, 517 (2008).

\bibitem{Osborne2002}
T. J. Osborne and M. A. Nielsen, Phys. Rev. A \textbf{66}, 032110 (2002).

\bibitem{Osterloh2002}
A. Osterloh, L. Amico, G. Falci, and R. Fazio, Nature \textbf{416}, 608 (2002).

\bibitem{GuEE}
S.J. Gu, H.Q. Lin, Y.Q. Li, Phys. Rev. A \textbf{68} 042330 (2003); S.J. Gu, G.S. Tian, H.Q. Lin, Phys. Rev. A \textbf{71} 052322 (2005); S.J. Gu, S.S. Deng, Y.Q. Li, H.Q. Lin, Phys. Rev. Lett. \textbf{93} 086402 (2004).

\bibitem{Levin2006}
M. Levin, X.G. Wen, Phys. Rev. Lett. \textbf{96} 110405 (2006).

\bibitem{Kitaev2006}
A. Kitaev, J. Preskill, Phys. Rev. Lett. \textbf{96} 110404 (2006).


\bibitem{Gu2013}
S. J. Gu, W. C. Yu, and H. Q. Lin,  Ann. Phys. \textbf{336}, 118 (2013).

\bibitem{YU2016} W.C. Yu, S.J. Gu and H.Q. Lin, Eur. Phys. J. B \textbf{89} 212 (2016).

\bibitem{Furukawa2006}
S. Furukawa, G. Misguich, and M. Oshikawa, Phys. Rev. Letts. \textbf{96}, 047211 (2006).

\bibitem{Henley2014}
C. L. Henley and H. J. Changlani, J. Stat. Mech. P11002 (2014).

\bibitem{Cheong2009}
S.-A. Cheong and C. L. Henley, Phys. Rev. B \textbf{79}, 212402 (2009).

\bibitem{ssh}
W.P. Su, J.R. Schrieffer, A.J. Heeger, Phys. Rev. Lett. {\bf 42}, 1698 (1979):
Phys. Rev. B {\bf 22}, 2099 (1980); A.J. Heeger, S- Kivelson, J.R. Schrieffer
and W.P. Su, Rev. Mod. Phys. {\bf 60}; 781 (1988).

\bibitem{hasan}
M. Z. Hasan and C. L. Kane,
 Rev. Mod. Phys. {\bf 82}, 3045 (2010)

\bibitem{szhang}
 X.-L. Qi and S.-C  Zhang,
 Rev. Mod. Phys.  {\bf 83}, 1057 (2011).

\bibitem{schnyder}
A.P. Schnyder, S. Ryu, A. Furusaki and A.W.W. Ludwig,
Phys. Rev. B {\bf 78}, 195125 (2008).

\bibitem{yakovenko}
S.S. Pershoguba and V.M. Yakovenko, Phys. Rev. B {\bf 86}, 075304 (2012).


\bibitem{white}
S.R. White et al., {\it Density-Matrix Renormalization: A new numerical method in physics} (Berlin:Springer, 1999).

\bibitem{wang}
C.L. Wang, W.Z. Wang, C.L. Gu, Z.B. Su and L. Yu, Phys. Rev. B {\bf 48}, 10788 (1993).

\bibitem{jeckelmann}
E. Jeckelmann, Phys. Rev. B {\bf 57}, 11838 (1998).

\bibitem{riera}
J. Riera and D. Poilblanc, Phys. Rev. B {\bf 62}, R16243 (2000).

\bibitem{zhang1}
Y.Z. Zhang, C.Q. Wu and H.Q. Lin, Phys. Rev. B {\bf 72}, 125126 (2005).

\bibitem{benthien}
H. Benthien, F.H.L. Essler and A. Grage, Phys. Rev. B {\bf 73}, 085105 (2006).

\bibitem{campbell}
D.K. Campbell, J. Tinka Gammel and E.Y. Loh, Jr., Phys. Rev. B {\bf 42}, 475 (1990).

\bibitem{ejima}
S. Ejima, F. Gebhard and S. Nishimoto, Phys. Rev. B {\bf 74}, 245110 (2006).

\bibitem{kumar}
M. Kumar, S. Ramasesha and Z.G. Soos, Phys. Rev. B {\bf 79}, 035102 (2009).

\bibitem{weber}
M. Weber, F.F. Assaad and M. Hohenadler, Phys. Rev. B {\bf 91}, 245147 (2015).

\bibitem{nakamura}
M. Nakamura, J. Phys. Soc. Jpn. {\bf 68}, 3123 (1999); Phys. Rev. B {\bf 61}, 16377 (2000).

\bibitem{sengupta}
P. Sengupta, A.W. Sandvik and D.K. Campbell, Phys. Rev. B {\bf 65}, 155113 (2002).

\bibitem{zhang2}
Y.Z. Zhang, Phys. Rev. Lett. {\bf 92}, 246404 (2004)

\bibitem{ejima2}
S. Ejima and S. Nishimoto, Phys. Rev. Lett. {\bf 99}, 216403 (2007).


\bibitem{MMWolf}
M. M. Wolf, F. Verstraete, M. B. Hastings, and J. I. Cirac, Phys.
Rev. Lett. \textbf{100}, 070502 (2008).

\bibitem{SJGuJPA}
S. J. Gu, C. P. Sun, and H. Q. Lin, J. Phys. A: Math. Theor. \textbf{41}, 025002 (2008).

\bibitem{kitaev}
A. Y. Kitaev, Phys.-Usp. 44, 131 (2001).



\bibitem{manmana}
S.R. Manmana, A.M. Essin, R.M. Noack and V. Gurarie, Phys. Rev. B {\bf 86}, 205119 (2012).

\bibitem{berry1}
H. Guo and S.Q. Shen, Phys. Rev. B {\bf 84}, 195107 (2011).

\bibitem{berry2}
H. Guo, S.Q. Shen and S. Feng, Phys. Rev. B {\bf 86}, 085124 (2012).

\end{thebibliography}
\end{document}